\documentclass[preprint,prb]{revtex4}

\usepackage{graphicx}
\usepackage{dcolumn}
\usepackage{float}
\usepackage{amsmath}

\newcommand{\comment}[1]{}

\begin{document}

\title{\boldmath Intrinsic temperature-dependent evolutions in the electron-boson spectral density obtained from optical data \unboldmath}

\author{Jungseek Hwang}\email{jungseek@skku.edu}

\affiliation{Department of Physics, Sungkyunkwan University, Suwon, Gyeonggi-do 440-746, Republic of Korea}

\date{\today}


%
%
\begin{abstract}
We investigate temperature smearing effects on the electron-boson spectral density function ($I^2\chi(\omega)$) obtained from optical data using a maximum entropy inversion method. We start with two simple model input $I^2\chi(\omega)$, calculate the optical scattering rates at selected temperatures using the model input spectral density functions and a generalized Allen's formula, then extract back $I^2\chi(\omega)$ at each temperature from the calculated optical scattering rate using the maximum entropy method (MEM) which has been used for analysis of optical data of high-temperature superconductors including cuprates, and finally compare the resulting $I^2\chi(\omega)$ with the input ones. From this approach we find that the inversion process can recover the input $I^2\chi(\omega)$ almost perfectly when the quality of fits is good enough and also temperature smearing (or thermal broadening) effects appear in the $I^2\chi(\omega)$ when the quality of fits is not good enough. We found that the coupling constant and the logarithmically averaged frequency are robust to the temperature smearing effects and/or the quality of fits. We use these robust properties of the two quantities as criterions to check whether experimental data have intrinsic temperature-dependent evolutions or not. We carefully apply the MEM to two material systems (one optimally doped and the other underdoped cuprates) and conclude that the $I^2\chi(\omega)$ extracted from the optical data contain intrinsic temperature-dependent evolutions.
\\ \\

\noindent *Correspondence to [email: jungseek@skku.edu].

\end{abstract}

\maketitle

In strongly correlated materials including high-temperature superconductors the information of correlation between charge carriers appears in their inelastic scattering spectra. The interaction between charge carriers can be described by the electron-boson spectral density function, which can be described by a model of exchanging the force-mediating bosons between electrons. Here we denote the electron-boson spectral density function as $I^2\chi(\omega)$, where $I$ is the coupling constant between the boson and an electron and $\chi(\omega)$ is the energy spectrum of the boson. In superconducting materials the electron-boson spectral density function can play an important role for forming electron-electron Cooper pairs for the superconductivity. Therefore this electron-boson function has been known as the glue (spectral) function. $I^2\chi(\omega)$ and/or $\chi(\omega)$ can be exposed experimentally by various spectroscopic experimental techniques\cite{carbotte:2011}. The glue function is also called the Eliashberg function\cite{carbotte:1990}. In cuprate systems this electron-boson density function shows universal temperature and doping dependent properties\cite{carbotte:2011}. Particularly, optical spectroscopic technique plays a crucial role to expose the temperature and doping dependent properties of the glue function since this technique can be used to study all cuprate systems. Usually one extracts the electron-boson spectral density function from the optical scattering rate (or the imaginary part of the optical self-energy) which can be defined by an extended Drude model\cite{puchkov:1996,hwang:2004} using generalized Allen's fomulas\cite{allen:1971,mitrovic:1985,shulga:1991,marsiglio:1998,sharapov:2005,schachinger:2006}. The extracting processes can be categorized into two groups: model-dependent and model-independent\cite{marsiglio:1998,dordevic:2005,schachinger:2006} processes. Particularly, one of the model-independent processes incorporated with a maximum entropy method\cite{schachinger:2006} has been used widely since its introduction and, in principle, allows us to obtain the most probable electron-boson spectral density functions from the optical data. The model-independent process does not impose any restrictions on the shape of $I^2\chi(\omega)$ except for one that the quantity is positive. Using this process a lot of important temperature and doping dependent properties of $I^2\chi(\omega)$ have been exposed from optical data\cite{hwang:2007,yang:2009,hwang:2013,hwang:2014}; in these studies the authors have used approximate Shulga {\it et al.}\cite{shulga:1991} or Sharapov and Carbotte\cite{sharapov:2005} formulas. There also have been some other optical studies\cite{heumen:2009,heumen:2009a,heumen:2009b} which show less temperature and doping evolutions in the extracted glue (or $I^2\chi(\omega)$) functions; in these study the authors have obtained a histogram representation of the glue function using a least-squares process and a full expression\cite{shulga:1991,allen:2015} for the optical conductivity.

In this paper we investigated the temperature smearing effects which might be caused by the maximum entropy inversion process. This issue will be an important problem to tell whether the temperature dependent-evolutions in $I^2\chi(\omega)$ extracted using the maximum entropy inversion process are intrinsic or extrinsic. We started with two model $I^2\chi(\omega)$ (one consists of a single Gaussian peak and the other two identical (or double) Gaussian peaks), calculated the optical scattering rates at selected temperatures using Shulga {\it et al.} formula\cite{shulga:1991} which is an integral equation relating the electron-boson spectral density to the optical scattering rate, then applied the maximum entropy inversion process\cite{schachinger:2006} to extract $I^2\chi(\omega)$ from the calculated optical scattering rates, and finally compared the resulting $I^2\chi(\omega)$ at the selected temperatures with the input $I^2\chi(\omega)$ to check whether there are any temperature-dependent properties other than the temperature smearing. From this approach we confirmed that the temperature smearing (or thermal broadening) effects on the extracted $I^2\chi(\omega)$ is dependent of the quality of fits and found that two physical quantities (the coupling constant and the averaged frequency of $I^2\chi(\omega)$) are robust to the quality of fits. We also carefully reanalyze optical data of two (optimally and underdoped) Bi-based cuprates with different fitting qualities (optimally doped) and a different approach (underdoped) to see whether the temperature-dependent properties in the experimental spectra are intrinsic or come from merely the temperature smearing. From these studies we get to a conclusion that the temperature-dependent trends of the extracted $I^2\chi(\omega)$ from optical data using the maximum entropy method are clearly intrinsic even though there are some unavoidable temperature smearing effects.

\section*{Model calculations and results}

For our model calculations we used two model input electron-boson spectral density functions: one consists of a single Gaussian peak, i.e. $I^2\chi(\omega) =\frac{A_p}{\sqrt{2\pi} (d/2.35)}  \exp\Big{[}\!-\!\frac{(\omega-\omega_p)^2}{2(d/2.35)^2} \Big{]}$ where $A_p$ is the Gaussian peak area of 31 meV, $d$ is the width of 10 mev, and $\omega_p$ is the center frequency of 60 meV and the other consists of two identical Gaussian peaks, i.e.  $I^2\chi(\omega) =\frac{A_{p,1}}{\sqrt{2\pi} (d_1/2.35)}  \exp\Big{[}\!-\!\frac{(\omega-\omega_{p,1})^2}{2(d_1/2.35)^2} \Big{]}+\frac{A_{p,2}}{\sqrt{2\pi} (d_2/2.35)}  \exp\Big{[}\!-\!\frac{(\omega-\omega_{p,2})^2}{2(d_2/2.35)^2} \Big{]}$ where $A_{p,1}$ and $A_{p,2}$ are the areas of the two Gaussian peaks with the same value of 31 meV, $d_1$ and $d_2$ are the widths of the two peaks with the same value of 10 mev, and $\omega_{p,1}$ and $\omega_{p,2}$ are the center frequencies with 60 meV and 120 meV, respectively, as shown in lower frames of Fig. 1 and Fig. 3. We calculated the optical scattering rates ($1/\tau^{op}(\omega,T)$) at selected temperatures from 5 K to 300 K for the two input $I^2\chi(\omega)$ using Eq. (\ref{eq1}) in the Method section and taking $1/\tau_{imp} =$ 0. The calculated optical scattering rates are displayed in the upper frames of Fig. 1 and Fig. 3 for the single and double Gaussian $I^2\chi(\omega)$, respectively.

Then we extracted the electron-boson spectral density functions ($I^2\chi(\omega)$) from the calculated optical scattering rates ($1/\tau^{op}(\omega)$) using the maximum entropy method (MEM) in order to see any temperature smearing effects. We controlled the quality of fits with the misfit parameter $\sigma$ as an adjustable parameter (refer to Eq. (\ref{eqc}) in the Method section). In Fig. 1(a) and 1(c) we display, respectively, fits and extracted $I^2\chi(\omega)$ by using the same misfit parameter $\sigma$ = 0.10 for all selected temperatures. Even though the quality of fits is quite good for all temperatures the resulting $I^2\chi(\omega)$ show some temperature-dependencies; as the temperature increases the extracted $I^2\chi(\omega)$ becomes broader compared with the input $I^2\chi(\omega)$. We call this temperature-dependent trend as temperature smearing or thermal broadening. We note that the extracted $I^2\chi(\omega)$ at 5 K is almost the same as the input $I^2\chi(\omega)$. For temperature 300 K case we also fit the data with larger misfit parameters ($\sigma$ = 0.30, 0.50, and 1.00) to see the misfit-dependent behavior and observe that the broadening becomes larger as the misfit parameter increases. In Fig. 1(b) and 1(d) we also display, respectively, fits with smaller misfit parameters ($\sigma$ = 0.10, 0.05, 0.02, and 0.007 from low to high temperatures) and the extracted $I^2\chi(\omega)$ for the same selected temperatures. In order to recover the input $I^2\chi(\omega)$ completely from the calculated (or theoretical) optical scattering rates we have to use the smaller value of the misfit parameter ($\sigma$) for the higher temperature as shown in Fig. 1(d).

Now we added random noises to the calculated optical scattering rate at 300 K and analyzed the new optical scattering rates using the MEM in order to investigate noise effects on the extracted $I^2\chi(\omega)$. We added two different amplitudes of random noises: one is 1 meV and the other 5 meV. We display the new optical scattering rates included the random noises of amplitudes of 1 meV and 5 meV, respectively, in Fig. 2(a) and 2(b). We fitted the new optical scattering rates using the MEM with various misfit parameters, which are displayed in the figure. Fits are quite good for all misfit parameters. For each case of the random noise the misfit parameter seems to approach a limiting value; $\sigma \sim$ 0.53323 for the amplitude of 1 meV and $\sigma \sim$ 2.66647 for the amplitude of 5 meV. The limiting value seems to be related to the amplitude of the random noise; the higher noise amplitude gives the larger limiting misfit value. We note that for the case of no noise the limiting value seems to be zero (refer to Fig. 1). In Fig. 2(c) and 2(d) we display the extracted $I^2\chi(\omega)$ for the two different noise cases, respectively and the input $I^2\chi(\omega)$. For the both cases we obtained much sharper $I^2\chi(\omega)$ than the input $I^2\chi(\omega)$ with the misfit parameters near the limiting value, which seems to be absent for the case of no noise.

In Fig. 3 we display fits and extracted $I^2\chi(\omega)$ for the input double Gaussian $I^2\chi(\omega)$. In Fig. 3(a) we show the calculated optical scattering rates at selected temperatures using Eq. (\ref{eq1}) in the Method section and fits to the calculated scattering rates using the maximum entropy method (MEM) with the same misfit parameter ($\sigma$) of 0.1 for the all selected temperatures. All fits are quite good. In Fig. 3(c) we display the corresponding extracted $I^2\chi(\omega)$ at all selected temperatures which show strong temperature-dependencies; at 300 K the two peaks are merged into a broad single peak and at 200 K the two peaks are resolved but their positions are red- (the lower frequency peak) and blue- (the higher frequency peak) shifted. At 5 K the extracted $I^2\chi(\omega)$ is almost the same as the input $I^2\chi(\omega)$. We note that while only peak broadening occurs for the single Gaussian case both peak broadening and shifting occur for the double Gaussian case. But for the double Gaussian case the broadening seems to cause the shifting; the peak shifting is a secondary effect. As we can see in Fig. 3(b) and 3(d) when we make fits tighter (or with smaller misfit parameters: $\sigma$ = 0.10, 0.05, 0.02, 0.01, and 0.001 from low to high temperatures) to the data at high temperatures we are able to recover almost completely the input double (or two-peak) Gaussian $I^2\chi(\omega)$. It is worth noting that since the calculated (or theoretical) optical scattering rates do not contain any errors (or any noises) we can recover the input $I^2\chi(\omega)$ perfectly. However, in general, experimentally measured optical scattering rates always contain some background uncertainties and because of these uncertainties as we could see previously (refer to the discussion with Fig. 2) one still may be able to fit to the data and extract the correct $I^2\chi(\omega)$ but it does not seem to be easy to find the right misfit parameter to obtain the correct $I^2\chi(\omega)$.

Interestingly, we find that some physical quantities are quite robust to the temperature smearing and/or the quality of fits. Those quantities are the coupling constant ($\lambda$) and the logarithmically averaged frequency ($\omega_{ln}$) which can be calculated from the electron-boson spectral density function ($I^2\chi(\omega)$). The two quantities can be defined as follows:
\begin{eqnarray}\label{eq0a}
\lambda(T) &\equiv& 2\int_0^{\omega_c}\frac{I^2\chi(\Omega,T)}{\Omega}d\Omega \nonumber \\
\omega_{ln}(T) &\equiv& \exp\Big{[} \frac{2}{\lambda(T)} \int_0^{\omega_c} \frac{\ln{\Omega}\:\: I^2\chi(\Omega,T)}{\Omega} d\Omega\Big{]}
\end{eqnarray}
where $\omega_c$ is the cutoff frequency; for this study we take the frequency as 300 meV. We calculated these two quantities with the all extracted $I^2\chi(\omega)$ so far. In Fig. 4(a) we display the obtained coupling constants ($\lambda$) as functions of temperature for the two input $I^2\chi(\omega)$ and both loose and tight fits; the coupling constants for all four cases show almost no temperature-dependencies. In Fig. 4(b) we display the obtained averaged frequency ($\omega_{ln}$) of $I^2\chi(\omega)$ as functions of temperature also for the two input $I^2\chi(\omega)$ and both loose and tight fits. This quantity shows a quite small temperature-dependence for the loose fit; for the worst case the averaged frequency at 300 K is reduced around 3\% compared with the value at 5 K for the double Gaussian and loose fit case (the red triangle). In Fig. 4(c) we display the coupling constant as a function of the quality of fits for the optical scattering rate obtained from the single Gaussian $I^2\chi(\omega)$ at 300 K (refer to Fig. 1(c) and 1(d)); the coupling constant shows almost no temperature-dependence. Therefore we conclude that the coupling constant is a quite robust quantity to both the temperature smearing and the quality of fits. In Fig. 4(d) we display the averaged frequency as a function of the quality of fits for the same case of Fig. 4(c); this quantity shows more dependence of the quality of fits than the coupling constant but still the dependence is quite small. We note that the averaged frequency increases with decreasing the misfit parameter. In Fig. 4(e) and 4(g) we display the coupling constants as functions of the quality of fits for the optical scattering rates included the random noises with amplitudes of 1 meV and 5 meV before performing the MEM inversion process (refer to Fig. 2). In Fig. 4(f) and 4(h) we display the corresponding averaged frequencies as functions of the quality of fits. The both quantities show similar fitting-quality-dependencies regardless of the amplitudes of random noises. From these results we learned that these two quantities are quite robust to both the temperature smearing (or thermal broadening) and the quality of fits and can be important measures for checking whether there are intrinsic temperature-dependent evolutions in experimental optical data at various temperatures.

\section*{Application to real material systems and discussion}

Now we look up two Bi-based cuprate systems: one is an optimally doped Bi$_2$Sr$_2$CaCu$_2$O$_{8+\delta}$ (Bi2212) with $T_c =$ 96 K and the other an underdoped Bi2212 with $T_c$ = 69 K. We denoted them as, respectively, Bi2212-OPT96A and Bi2212-UD69. These two systems have been analyzed using a similar approach and the studies have been published already\cite{hwang:2007,hwang:2011}. However here we focus on the issue whether these material systems contain intrinsic temperature-dependent evolutions or their temperature-dependent properties are byproducts of the inversion process. To resolve the issue explicitly we reanalyzed Bi2212-OPT96A data with two different qualities of fits using the maximum entropy method and analyze Bi2212-UD69 data, {\it for the first time}, using the maximum entropy inversion process.

We applied the maximum entropy method to Bi2212-OPT96A at various temperatures with two different sets of the misfit parameters which are described below. We note that the larger set of the misfit parameters ($\sigma$) is similar to the one used in the published literature\cite{hwang:2007}. We compared the resulting $I^2\chi(\omega)$ obtained from the MEM applications each other to see any intrinsic temperature-dependent evolutions in the experimental data. We display the optical scattering rate data at normal states and fits in Fig. 5(a) and the corresponding extracted $I^2\chi(\omega)$ in Fig. 5(b). In the inset of Fig. 5(b) we show the peak positions as functions of temperature for the loose (or larger misfit parameters) and tight (or smaller misfit ones) fit cases; both sets of data show similar values of the peak positions and a similar temperature-dependent trend. The larger set of the misfit parameters ($\sigma$) are 3.5 at 102 K, 2.3 at 200 K, and 2.7 at 300 K for the loose fits and the corresponding smaller set of the parameters for the tight fits are 1.8, 1.28, and 1.55, respectively. We note that the extracted $I^2\chi(\omega)$ for the tight fits seem to be non-physical since for the three temperatures there are spectral gaps (or no spectral weights) in low frequency region below $\sim$40 meV for 102 K and $\sim$60 meV for 200 K and 300 K, which have not been observed. We also note that the sharper peaks of $I^2\chi(\omega)$ extracted with the smaller misfit parameters may be obtained because of experimental uncertainties (refer to Fig. 2(c) and 2(d)). We also display the two robust quantities discussed previously (refer to Fig. 4): the coupling constant in Fig. 5(c) and the logarithmically averaged frequency in Fig. 5(d). Both sets of the coupling constant obtained by different qualities of fits show similar temperature-dependencies and the set from the tight fits has slightly lower values. We emphasize that two sets of the coupling constants show clear strong temperature-dependencies regardless of the qualities of fits. These temperature-dependencies are too large ($\sim$23\% decrease in $\lambda$ from 101 K to 300 K) to be caused by the temperature smearing or the quality of fits if we consider less than 1\% decrease in $\lambda$ from 5 K to 300 K for single Gaussian case (refer to Fig. 4(a)). The two sets of the averaged frequencies obtained using different qualities of fits are quite different. The set obtained by the tight fits shows smaller values than those by the loose fits; this is opposite to the fitting-quality-dependent trend of the averaged frequency (refer to Fig. 4(d)). However the temperature-dependent trends of the two sets of the averaged frequencies are similar to each other. We expect that the large difference and the opposite trend may come from unknown experimental uncertainties which every experimental data may have. Our results indicate that the optical data of Bi2212-OPT96A sample contain intrinsic temperature-dependent evolutions even though they show some dependencies on the quality of fits.

Now we applied the maximum entropy method (MEM) to optical data of the underdoped Bi2212-UD69 sample. The electron-boson spectral density functions ($I^2\chi(\omega)$) of this material have been extracted\cite{hwang:2011}. But in the previous study the author modeled the shape of $I^2\chi(\omega)$ with two (sharp and broad) components and fitted the data with a least-squares process. Here we do not give any constraints on the shape of $I^2\chi(\omega)$ except for a requirement that the quantity is positive. We need to use the generalized kernel Eq. (\ref{eq2})\cite{sharapov:2005} in the Method section for the Allen's formula to analyze this underdoped cuprate since we have to take care of the pseudogap\cite{timusk:1999} of the underdoped Bi2212-UD69 sample. We adopted the same shape of the pseudogap which has been used previously\cite{hwang:2008,hwang:2011}; in this pseudogap model the density of states loss in the pseudogap is recovered just above the pseudogap. The symmetrized and normalized density of states, $\tilde{N}(\omega)$ (or the pseudogap), can be described as follows:
\begin{eqnarray}\label{eq0b}
\tilde{N}(\omega, T) \!\!\!&=&\!\!\! \tilde{N}(0, T) \!+ \![1\!-\!\tilde{N}(0, T)]\Big{(}\frac{\omega}{\Delta_{PG}} \Big{)}^2 \:\: \mbox{for} \:\:|\omega|\! \leq\! \Delta_{PG}, \nonumber \\
&=& 1 + \frac{2}{3}[1-\tilde{N}(0, T)] \:\:\:\mbox{for} \:\:|\omega| \!\in \!(\Delta_{PG}, 2\Delta_{PG}), \nonumber \\
&=& 1 \:\:\: \:\:\:\:\:\: \:\:\:\:\:\: \:\:\:\mbox{for}  \:\:\:|\omega| > 2\Delta_{PG}
\end{eqnarray}
where $\Delta_{PG}$ is the size of the pseudogap and $\tilde{N}(0, T)$ is the density of states at the Fermi energy (or zero frequency). We note that $1-\tilde{N}(0, T)$ is a measure of the strength (or depth) of the pseudogap. We used the temperature-dependent model $\tilde{N}(0,T)$ observed by Kanigel {\it et al.}\cite{kanigel:2006,hwang:2008b}, i.e. $\tilde{N}(0,T) \simeq  0.67 \:T/T^*$ for $T \leq T^*$ and 1.0 for $T > T^*$, where $T^*$ is the pseudogap (onset) temperature. For this analysis we take $T^* = $ 300 K and $\Delta_{PG} =$ 43.3 meV. In Fig. 6(a) we display the optical scattering rate data and fits using the maximum entropy method at various temperatures of normal states. The misfit parameters ($\sigma$) for these fits are, respectively, 4.2, 3.1, 3.8, 3.3, 3.4, 3.4, 3.5, and 3.7 from low to high temperatures. We needed to have the impurity scattering rates to remove non-physical upturns in low frequency region\cite{hwang:2013}. The impurity scattering rates ($1/\tau_{imp}$) are 10, 10, 0, 0, 0, 30, 80, and 80 meV, respectively, from low to high temperatures. In Fig. 6(b) we display the extracted $I^2\chi(\omega)$ at various temperatures which have a dominant single peak and show strong temperature-dependencies; the thermal broadening may exist in the extracted $I^2\chi(\omega)$ and the peak position clearly shifts to higher frequency with increasing temperature. But if we consider only the temperature smearing this peak-shift is not expected (refer to Fig. 1 and related discussion). These results look similar to the reported $I^2\chi(\omega)$\cite{hwang:2011} but as we pointed out previously in this new work the shape of $I^2\chi(\omega)$ is not modeled. In the inset we show the temperature-dependent evolution of the peak position in the extracted $I^2\chi(\omega)$; this temperature-dependent trend of the peak position is slightly different from that in the reported literature\cite{hwang:2011}. This difference can be attributed to the different constraints on the shape of $I^2\chi(\omega)$ in the two different analysis methods (one is the MEM and the other a least-squared method). The peak position shows an anomaly (a kink) near 150 K; above the temperature the peak position decreases almost linearly with decreasing temperature and below the temperature the position seems to be fixed at around 26 meV. This characteristic temperature might be related to the onset temperature of the magnetic resonance mode which was observed by inelastic neutron scattering experiments\cite{rossat:1991,dai:1999}. In Fig. 6(c) and 6(d) we display, respectively, the coupling constant ($\lambda$) and the averaged frequency ($\omega_{ln}$) as functions of temperature. Both quantities show strong temperature-dependencies; while the coupling constant increases, as lowering temperature, almost linearly from 1.8 at 295 K to 5.7 at 70 K the average frequency decreases with reducing temperature and shows a kink near 200 K. These strong temperature-dependencies ($\sim$68\% decrease in $\lambda$ from 70 K to 295 K) cannot be explained with the temperature smearing effect (less than 1\% decrease in $\lambda$ from 5 K to 300 K for the single Gaussian case) which can be caused by the maximum entropy inversion process; if we consider the results of our previous model calculations the temperature-dependencies observed in the experimental data are too large to be caused by the temperature smearing effect. These results also indicate that the experimental data of Bi2212-UD69 sample clearly contain intrinsic temperature-dependent evolutions. We note that similar strong temperature-dependent results have been obtained by Hwang\cite{hwang:2011} using a least-squares fit analysis of the same material system.

\section*{Comparison of the approximate and full expressions for the optical conductivity}

So far we have used the approximate formulas\cite{shulga:1991,sharapov:2005} (Eq. (\ref{eq1}) and Eq. (\ref{eq2})) to produce the theoretical data and to analyze both the theoretical and experimental data using the maximum entropy inversion process. Therefore one question which one may ask would be that if the full expression (Eq. (\ref{eq3})) for the conductivity\cite{shulga:1991,allen:2015} (refer to the Method section) instead of the approximate formulas\cite{shulga:1991,sharapov:2005} is applied, are the results and conclusion obtained previously still maintained? To answer this question we performed the following study. First we compare the optical scattering rates obtained using both approximate and full expressions at various temperatures for the two model input (single and double Gaussian) $I^2\chi(\omega)$ cases. The resulting optical scattering rates for the two single and double Gaussian $I^2\chi(\omega)$ cases are displayed in Fig. 7(a) and 7(b), respectively. At low temperatures below 100 K the two optical scattering rates obtained using two different formulas (Eq. (\ref{eq1}) and Eq. (\ref{eq3})) agree each other quite well in a wide spectral range. At 100 K these results are similar to those of Shulga {\it et al}\cite{shulga:1991}. As higher temperatures the two scattering rates show significant discrepancies in low frequency region below 100 meV and the discrepancy becomes larger as temperature increases.

Now we extract the electron-boson spectral density function ($I^2\chi(\omega)$) from the optical scattering rates obtained with the full expression using the approximate formula and the maximum entropy inversion process to see any serious differences in the temperature-dependent properties between two optical scattering rates obtained with the two different formulas. The resulting fits, the data, and extracted $I^2\chi(\omega)$ are displayed in Fig. 8 (a)-(d) for both single and double input Gaussian $I^2\chi(\omega)$ cases. For the single Gaussian case the fitting quality become worse as the temperature increases and extracted $I^2\chi(\omega)$ shows a single peak located at a similar peak frequency of the input $I^2\chi(\omega)$. We also calculated the coupling constant ($\lambda$) and the logarithmically averaged frequency ($\omega_{ln}$) (refer to Eq. (\ref{eq0a})) from the extracted $I^2\chi(\omega)$ and displayed them as functions of temperature in the insets of Fig. 8(b) and 8(d), respectively. Interestingly, both quantities show small temperature-dependencies: $\pm$0.3\% of the average 1.02 for $\lambda$ and $\pm$0.7\% of the average 62.48 meV for $\omega_{ln}$. The absolute values ($\lambda \simeq$ 1.02 and $\omega_{ln} \simeq$ 62.48 meV) are similar to those ($\lambda \simeq$ 1.05 and $\omega_{ln} \simeq$ 59.25 meV) in Fig. 4(a) and 4(b); while the coupling constants show $\sim$3\% lower than those in Fig. 4(a) the averaged frequencies show $\sim$5\% higher than those in Fig. 4(b). For the double Gaussian case the fitting qualities for all temperatures are quite good. But the extracted $I^2\chi(\omega)$ functions show some discrepancies compared with the input $I^2\chi(\omega)$, particularly for 300 K; at this temperature the two peaks are not resolved well. Interestingly the coupling constant and averaged frequency still show small temperature dependencies ($\pm$0.2\% of the average 1.51 for $\lambda$ and $\pm$2.6\% of the average 78.18 for $\omega_{ln}$) even though the absolute values ($\lambda \simeq$ 1.51 and $\omega_{ln} \simeq$ 78.18 meV) are slight different from those ($\lambda \simeq$ 1.68 and $\omega_{ln} \simeq$ 76.40 meV) in Fig. 4(a) and 4(b). While the coupling constants show $\sim$10\% lower than those in Fig. 4(a) the averaged frequencies show $\sim$2\% higher than those in Fig. 4(b). This study allows us to get a conclusion that the two robust quantities obtained using the maximum entropy inversion process with either approximate or full formula show small temperature-dependencies even though their absolute values may be slightly different from the real ones. In other word, we expect that application of either formula to measured experimental data will lead to the same conclusion as long as we consider the temperature-dependent intrinsic properties.

\section*{Conclusion}

We investigated an issue whether there are any intrinsic temperature-dependent trends in $I^2\chi(\omega)$ extracted from measured optical scattering rates using the maximum entropy inversion process. From model calculations we learned that temperature smearing (or thermal broadening) in the extracted $I^2\chi(\omega)$ might occur when the quality of fits was not good enough. This temperature smearing might cause peak-shifts (or spectral weight redistributions) for the input $I^2\chi(\omega)$ which consists of two identical (or double) Gaussian peaks. We also found that the coupling constant ($\lambda$) and the logarithmically averaged frequency ($\omega_{ln}$) are quite robust to the quality of fits and these quantities can be used to judge existence of intrinsic temperature-dependent properties in the extracted $I^2\chi(\omega)$. These two quantities have also important physical meanings: the coupling constant shows the intensity of the electron-electron interaction by exchanging the mediated bosons and the averaged frequency is closely related to the superconducting transition temperature which can be estimated by the generalized McMillan formula\cite{mcmillan:1968,williams:1989}. We revisited two Bi-based cuprate systems (Bi2212-OPT96A and Bi2212-UD69) to see intrinsic temperature-dependent evolutions in the extracted $I^2\chi(\omega)$ using the maximum entropy method. From these studies we conclude that these two cuprate systems have intrinsic temperature-dependent evolutions since the coupling constant and the averaged frequency show strong temperature-dependencies which cannot be explained by the temperature smearing effect.  We hope that our findings attract attentions from researchers in the field of superconductivity and make a step forward for figuring out the nature of the Cooper-paring glue of the high-temperature superconductors.

\newpage

\section*{Methods}

\section*{Analysis formalisms}
Allen has derived an integral equation which relates linearly the electron-boson spectral density function ($I^2\chi(\omega)$) to the optical scattering rate ($1/\tau^{op}(\omega)$) or the imaginary part of the optical self-energy ($\tilde{\Sigma}^{op}(\omega) \equiv \Sigma^{op}_1(\omega)+i\Sigma^{op}_2(\omega)$) for both normal and superconducting states\cite{allen:1971}. The Allen's original formulas can be used only for $T =$ 0 K and a constant density of states. A generalized formula, which can be used for finite temperature and normal state with a constant density of states, has been derived by Shulga {\it et al.}\cite{shulga:1991}. The Shulga {\it et al.}'s formula can be written as follows:
\begin{eqnarray}\label{eq1}
\frac{1}{\tau^{op}(\omega, T)} &=& \int_0^{\infty}d\Omega \:I^2\chi(\Omega, T) K(\omega,\Omega,T) + \frac{1}{\tau_{imp}} \:\:\mbox{,} \nonumber \\ K(\omega,\Omega,T)\!\! &=& \!\!2\omega \coth\Big{(}\frac{\Omega}{2T}\Big{)}\!-\!(\omega+\Omega)\coth\Big{(}\frac{\omega+\Omega}{2T}\Big{)}
\nonumber \\ \!\!&+&\!\!\!(\omega-\Omega)\coth\Big{(}\frac{\omega-\Omega}{2T}\Big{)},
\end{eqnarray}
where $1/\tau^{op}(\omega,T)$ is the optical scattering rate which can be related to the imaginary part of the optical self-energy as  $1/\tau^{op}(\omega,T) \equiv -2\Sigma^{op}_2(\omega, T)$, $I^2\chi(\omega,T)$ is the electron-boson spectral density function, and $1/\tau_{imp}$ is the impurity scattering. We note that $K(\omega,\Omega,T)$ is the Shulga {\it et al.}'s kernel which contains the temperature factor\cite{shulga:1991}. We used Eq. (\ref{eq1}) to obtain the optical scattering rates at selected temperatures from the input electron-boson spectral density functions and also to extract electron-boson spectral density functions from the calculated optical scattering rates using a maximum entropy method\cite{jaynes:1957,sivia:1990,schachinger:2006}.

\section*{Generalized kernel for the Allen's formula}
In order to analyze underdoped cuprates, which have the intriguing pseudogaps\cite{timusk:1999}, one needs to include the pseudogap (or non-constant density of states) in the model. A generalized Allen's formula, which can take care of the pseudogaps, was derived by Sharapov and Carbtte\cite{sharapov:2005}. The kernel of the generalized Allen's formula can be written as follows:
\begin{eqnarray}\label{eq2}
K(\omega,\Omega,T)\!\! &=& \!\!\frac{\pi}{\omega}\int^{+\infty}_{-\infty}\!\!d\nu \tilde{N}(\nu\!-\!\Omega)[n_B(\Omega)\!+\!1\!-\!n_F(\nu\!-\!\Omega)] \nonumber \\
\!\! &\times& \!\! [n_F(\nu\!-\!\omega)\!-\!n_F(\nu\!+\!\omega)]
\end{eqnarray}
where $n_B$ and $n_F$ are the Bose-Einstein and Fermi-Dirac distribution functions, respectively, which take care of the temperature dependencies, and $\tilde{N}(\omega)$ is the symmetrized electronic density of states ($N(\omega)$), i.e. $\tilde{N}(\omega)\equiv[N(\omega)+N(-\omega)]/2$, which takes care of any energy-dependencies in the density of states including the pseudogaps. For extracting $I^2\chi(\omega,T)$ of the underdoped cuprate from measured $1/\tau^{op}(\omega,T)$ we used the kernel of Eq. (\ref{eq2}) and a maximum entropy method\cite{jaynes:1957,sivia:1990,schachinger:2006}. 

\section*{Full expression for the optical conductivity}
The previous formulas (Eq. (\ref{eq1}) and Eq. (\ref{eq2})) derived by Shulga's {\it et al.} and Sharapov and Carbotte are approximate\cite{shulga:1991,sharapov:2005}; in general, they are valid at high frequencies, i.e. $1/[\omega \tau^{op}(\omega)] \ll 1$. However, Shulga {\it et al.} show that these formulas are valid in a wider frequency range at 100 K\cite{shulga:1991}. The full (non-approximate) expression\cite{shulga:1991,allen:2015} for the optical conductivity can be written as follows:
\begin{eqnarray}\label{eq3}
 \tilde{\sigma}(\omega,T) &=& i \frac{\omega_p^2}{4 \pi \omega} \int^{+\infty}_{-\infty} dx \frac{n_F(x)-n_F(x+\omega)}{\omega - \tilde{\Sigma}(x+\omega,T)+\tilde{\Sigma}^*(x,T)+i(1/\tau_{imp})} \\ \nonumber
 &=& i\frac{\omega_p^2}{4 \pi \omega} \frac{1}{1+[-2\tilde{\Sigma}^{op}(\omega,T)]/\omega},
\end{eqnarray}
where $\tilde{\sigma}(\omega)$ is the complex optical conductivity, $\omega_p$ is the plasma frequency, $1/\tau_{imp}$ is the impurity scattering rate, $\tilde{\Sigma}^*(x)$ is a complex conjugate of $\tilde{\Sigma}(x)$, and $\tilde{\Sigma}(x,T)$ is the single particle self-energy, which can be written as follows\cite{mahan,allen:1982}:
\begin{eqnarray}\label{eq4}
\tilde{\Sigma}(x, T) &=& \int^{\infty}_{0} d\Omega \:I^2\chi(\Omega) \int^{+\infty}_{-\infty}d\epsilon \Big{[} \frac{n_B(\Omega)+n_F(\epsilon)}{x-\epsilon+\Omega + i\delta} + \frac{n_B(\Omega)+1-n_F(\epsilon)}{x-\epsilon-\Omega - i\delta} \Big{]} \\ \nonumber
&=& \int^{\infty}_{0} d\Omega \: I^2\chi(\Omega)\Big{[} \psi\Big{(} \frac{1}{2}+i\frac{x-\Omega}{2 \pi T} \Big{)} -\psi\Big{(} \frac{1}{2}+i\frac{x+\Omega}{2 \pi T} \Big{)} -i \pi \coth\Big{(} \frac{\Omega}{2 T} \Big{)} \Big{]},
\end{eqnarray}
where $\psi(x)$ is the digamma function.

\section*{Maximum entropy method}
We used the maximum entropy inversion process introduced by Schachinger {\it et al.}\cite{schachinger:2006} to extract the electron-boson spectral density function from the optical scattering rate. We briefly introduce the maximum entropy method here. The maximum entropy method is based on the Bayes' theorem which provides the only consistent bridge between indirect (or posterior) and direct (or likelihood) probabilities\cite{sivia:1990}. The theorem can be described as follows:
\begin{equation}\label{eqa}
P(H|D,\aleph) = \frac{P(D|H,\aleph)P(H|\aleph)}{P(D|\aleph)},
\end{equation}
where $H$ stands for the hypothesis which we wish to infer, $D$ means the data, and $\aleph$ is any prior knowledge (or available background information), which can be the theoretically modeled kernel $K(\omega,\Omega,T)$ and any experimental sources of uncertainty\cite{schachinger:2006}. $P(H|D,\aleph)$ is the posterior probability distribution function (pdf), $P(D|H,\aleph)$ is the likelihood pdf, $P(H|\aleph)$ is the prior pdf, and $1/P(D|\aleph)$ is a normalization factor. In the maximum entropy method the appropriate prior for a positive and additive distribution can be of a special form as
\begin{equation}\label{eqa1}
P(H[f]|\aleph,\alpha,m) \propto \exp[\alpha S(f,m)],
\end{equation}
where $H[f]$ is the hypothesis functional of $f$, $f$ is positive and additive in our case, and $\alpha$ is a dimensionless parameter (initially unknown). $S$ is the generalized Shannon-Jayes entropy which can be written as follows:
\begin{equation}\label{eqb}
S(f,m) = \int\Big{[} f(x) - m(x) - f(x) \ln\Big{|} \frac{f(x)}{m(x)} \Big{|}\Big{]}dx,
\end{equation}
where $f(x)$ needs to be estimated with the highest probability through the maximum entropy process, $m(x)$ is a default model, which is usually taken to be a constant. In principle, data are independent each other and are subject to additive Gaussian noise. Then the likelihood pdf can be written as
\begin{equation}\label{eqb1}
P(D|H[f],\aleph) \propto \exp(-[\chi(f)]^2/2),
\end{equation}
where $\chi^2$ is the misfit, which measures how well a trial (or hypothesis) functional of $f$ (or $H[f]$) fits to the data ($D$). The $[\chi(f)]^2$ can be written as follows:
\begin{equation}\label{eqc}
[\chi(f)]^2 = \sum_{k=1}^{N} \frac{(D_k-H_k[f])^2}{\sigma^2},
\end{equation}
where $N$ is the number of data and $\sigma$ is an adjustable input parameter which is a measure of fitting quality. We call the $\sigma$ as the misfit parameter. Then using Eq. (\ref{eqa1}) and Eq. (\ref{eqb1}) the posterior pdf (or Eq. (\ref{eqa})) for $f$ can be written as follows:
\begin{equation}\label{eqd}
P(H[f]|D,\aleph,\alpha,m) = \frac{\exp\{\alpha \:S(f,m) - [\chi(f)]^2/2\}}{P(D|\aleph,m)}.
\end{equation}
Then the distribution which maximizes this posterior pdf (i.e. $\alpha\: S(f,m) - [\chi(f)]^2/2$) through a general algorithm provided by Skilling {\it et al.}\cite{skilling:1984} will give best estimate of $f$. In our case the data is the optical scattering rate, i.e. $D_k = 1/\tau^{op}(\omega_k)$, the hypothesis is the calculated optical scattering rate using a trial function $f(\omega_i) = I^2\chi(\omega_i)$, i.e. $H_k = 1/\tau^{op}_{hypothesis}(\omega_k) = \sum_{j=0}^{\infty} K(\omega_k,\Omega_j) I^2\chi(\Omega_j) \Delta\Omega$, and $m(\omega_i)$ is initially a constant, which means that we do not impose any particular structure (or shape) to the initial input $I^2\chi(\omega)$. But this initial condition may cause broadening in resulting $I^2\chi(\omega)$. In our maximum entropy process we iterated the process with an input value of the misfit parameter ($\sigma$) until we reach a criterion\cite{sivia:1990,schachinger:2006}, $\chi^2 = N$, where $N$ is the number of data, then the $\alpha$ parameter is determined automatically. Eventually, we extracted the most probable $f(\omega) = I^2\chi(\omega)$ and calculated the corresponding hypothesis (or the fit) $1/\tau^{op}_{hypothesis}(\omega)$ under the given condition ($\sigma$) for the optical scattering rate data ($1/\tau^{op}(\omega)$) at each temperature.

One can obtain smoothness of trial function $f(x) = I^2\chi(x)$ by introducing a hidden image, $h(x)$, which is blurred by a Gaussian as follows\cite{schachinger:2006}:
\begin{equation}\label{eqf}
f(x_i) = \sum_k B_{ik} h(x_k), \:\:\:\:\: B_{ik} = \frac{1}{2 \pi b^2} \exp\Big{[} -\frac{(x_i - x_k)^2}{2 b^2} \Big{]},
\end{equation}
where $b$ is the blur-width which is a hyperparameter and can be determined simultaneously with $\alpha$ by maximizing $P(H[f,h]|D,\aleph,\alpha,m)$. Here $f(x)$ gets into the likelihood pdf (Eq. (\ref{eqb1})) while $h(x)$ gets into the entropy $S$ (Eq. (\ref{eqb})). For our maximum entropy process we do not apply a blur value i.e. $b =$ 0, which means that all positive discrete $f(x_i)$ can be realized as $f(x_i) = h(x_i)$. We note that the interval between two consecutive discrete frequency variables is 1.0 meV.

%
%

\bibliographystyle{naturemag}
\bibliography{bib}

\newpage

\begin{figure}[!htbp]
\caption{The calculated optical scattering rates using the Shulga's formula, Eq. (\ref{eq1}) in the Method section and fits (frames (a) and (b)) at selected temperatures for the input single sharp Gaussian $I^2\chi(\omega)$ as shown in frames (c) and (d). In frames (c) and (d) we also show the extracted $I^2\chi(\omega)$ for loose and tight fits, respectively, using the maximum entropy method.}
 \label{fig1}
\end{figure}

\begin{figure}[!htbp]
\caption{The calculated optical scattering rates included random noises of amplitudes of 1 meV and 5 meV at 300 K are displayed, respectively, in frames (a) and (b) along with their fits obtained using the maximum entropy method with various misfit parameters, $\sigma$. Frames (c) and (d) display, respectively, the corresponding extracted $I^2\chi(\omega)$ for the two different amplitudes of the random noises with 1 meV and 5 meV, respectively.}
 \label{fig2a}
\end{figure}

\begin{figure}[!htbp]
\caption{The calculated optical scattering rates using the Shulga's formula, Eq. (\ref{eq1}) in the Method section and fits (frames (a) and (b)) at selected temperatures for the input two sharp Gaussian $I^2\chi(\omega)$ as shown in frames (c) and (d). In frames (c) and (d) we also show the extracted $I^2\chi(\omega)$ for loose and tight fits, respectively, using the maximum entropy method.}
 \label{fig2}
\end{figure}

\begin{figure}[!htbp]
\caption{Frames (a) and (b) show, respectively, the coupling constant ($\lambda(T)$) and the logarithmically averaged frequency ($\omega_{ln}(T)$) of the extracted $I^2\chi(\omega)$ from the loose and tight fits for the input single and double Gaussian peaks. Frames (c) and (d) show, respectively, the coupling constant ($\lambda(\sigma)$) and the averaged frequency ($\omega_{ln}(\sigma)$) of the extracted $I^2\chi(\omega)$ using five different qualities ($\sigma$) of fits for the input single Gaussian peak at 300 K. Frames (e) and (f) display, respectively, the coupling constant ($\lambda$) and the averaged frequency ($\omega_{ln}$) of the extracted $I^2\chi(\omega)$ with seven misfit parameters ($\sigma$) from the optical scattering rate included the random noises of amplitude of 1 meV for the input single Gaussian peak at 300 K. Frames (g) and (h) display the same quantities as in frames (e) and (f) except for a larger amplitude of random noise with 5 meV.}
 \label{fig3}
\end{figure}

\begin{figure}[!htbp]
\caption{Frame (a) shows optical scattering rate data of an optimally doped Bi2212 (Bi2212-OPT96A) and fits using the maximum entropy method (MEM) at three temperatures above $T_c$, 96 K. Frame (b) shows extracted electron-boson spectral density functions ($I^2\chi(\omega)$) using the MEM fits with two different misfit parameters at each temperature (see in the text). In the inset we compare two sets of temperature-dependent peak positions of the $I^2\chi(\omega)$ obtained with two different misfit parameters. In frames (c) and (d) we display, respectively, temperature-dependent coupling constant ($\lambda$) and logarithmically averaged frequency ($\omega_{ln}$) of extracted $I^2\chi(\omega)$ with the two different misfit parameters.}
 \label{fig4}
\end{figure}

\begin{figure}[!htb]
\caption{Frame (a) shows the optical scattering rates of underdoped Bi2212 (Bi2212-UD69) and fits using the maximum entropy method (MEM) at eight different temperatures above $T_c$, 69 K. Frame (b) shows the extracted electron-boson spectral density functions ($I^2\chi(\omega)$) using the MEM inversion process. In the inset we display temperature-dependent peak position of the extracted $I^2\chi(\omega)$. In frames (c) and (d) we display, respectively, temperature-dependent coupling constant ($\lambda$) and logarithmically averaged frequency ($\omega_{ln}$) of the extracted $I^2\chi(\omega)$.}
 \label{fig5}
\end{figure}

\begin{figure}[!htb]
\caption{The optical scattering rates obtained using two different formulas: one is the approximate Shulga {\it et al.} formula and the other the full expression, Eq. (\ref{eq1}) and Eq. (\ref{eq3}), respectively. Frames (a) and (b) show the calculated optical scattering rates at four different temperatures for the single Gaussian case and double Gaussian cases of input $I^2\chi(\omega)$, respectively.}
 \label{fig6}
\end{figure}

\begin{figure}[!htb]
\caption{The optical scattering rates obtained with the full expression (Eq. (\ref{eq3})) at various temperatures and their corresponding fits using the approximate formula (Eq. (\ref{eq1})) and the maximum entropy inversion process. Frames (a) and (c) show data and resulting fits for the single and double Gaussian cases of input $I^2\chi(\omega)$, respectively. Frames (b) and (d) display extracted corresponding $I^2\chi(\omega)$ using the maximum entropy process for the two cases of input $I^2\chi(\omega)$.}
 \label{fig7}
\end{figure}

\newpage

\noindent {\bf Author Contribution} JH done all work for the manuscript.
\\ \\

\noindent {\bf Acknowledgements} JH acknowledges financial support from the National Research Foundation of Korea (NRFK Grant No. 2013R1A2A2A01067629).
\\ \\

\noindent {\bf Competing Interests} The authors declare that they have no competing financial interests.
\\ \\

\noindent {\bf Correspondence} Correspondence and requests for materials should be addressed to Jungseek Hwang~(email: jungseek@skku.edu).

\newpage

\begin{figure}[t]
  \vspace*{-0.3 cm}%
  \centerline{\includegraphics[width= 7.0 in]{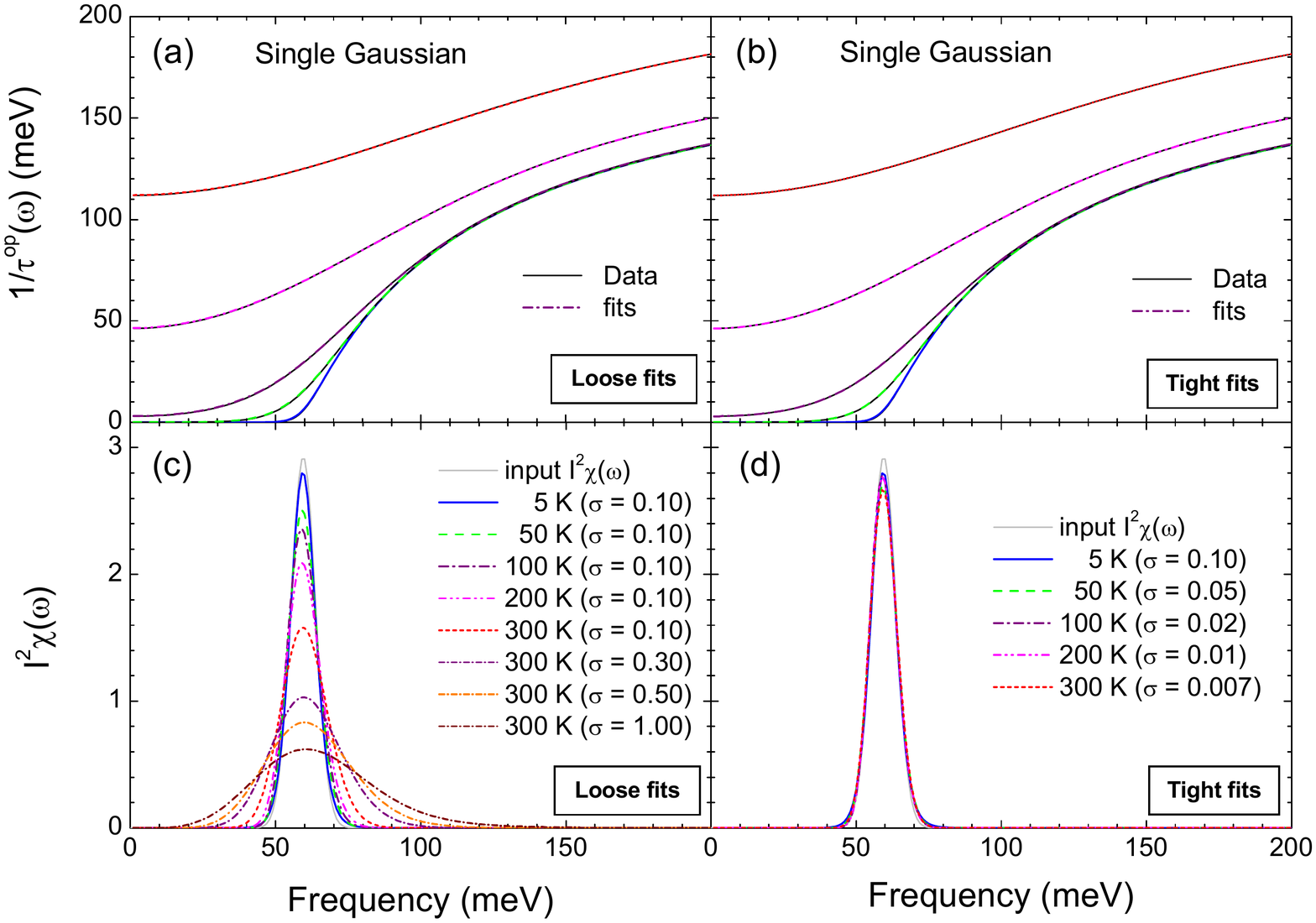}}%
  \vspace*{-0.6cm}%
 \label{fig1}
\end{figure}

\begin{figure}[t]
  \vspace*{-0.3 cm}%
  \centerline{\includegraphics[width= 7.0 in]{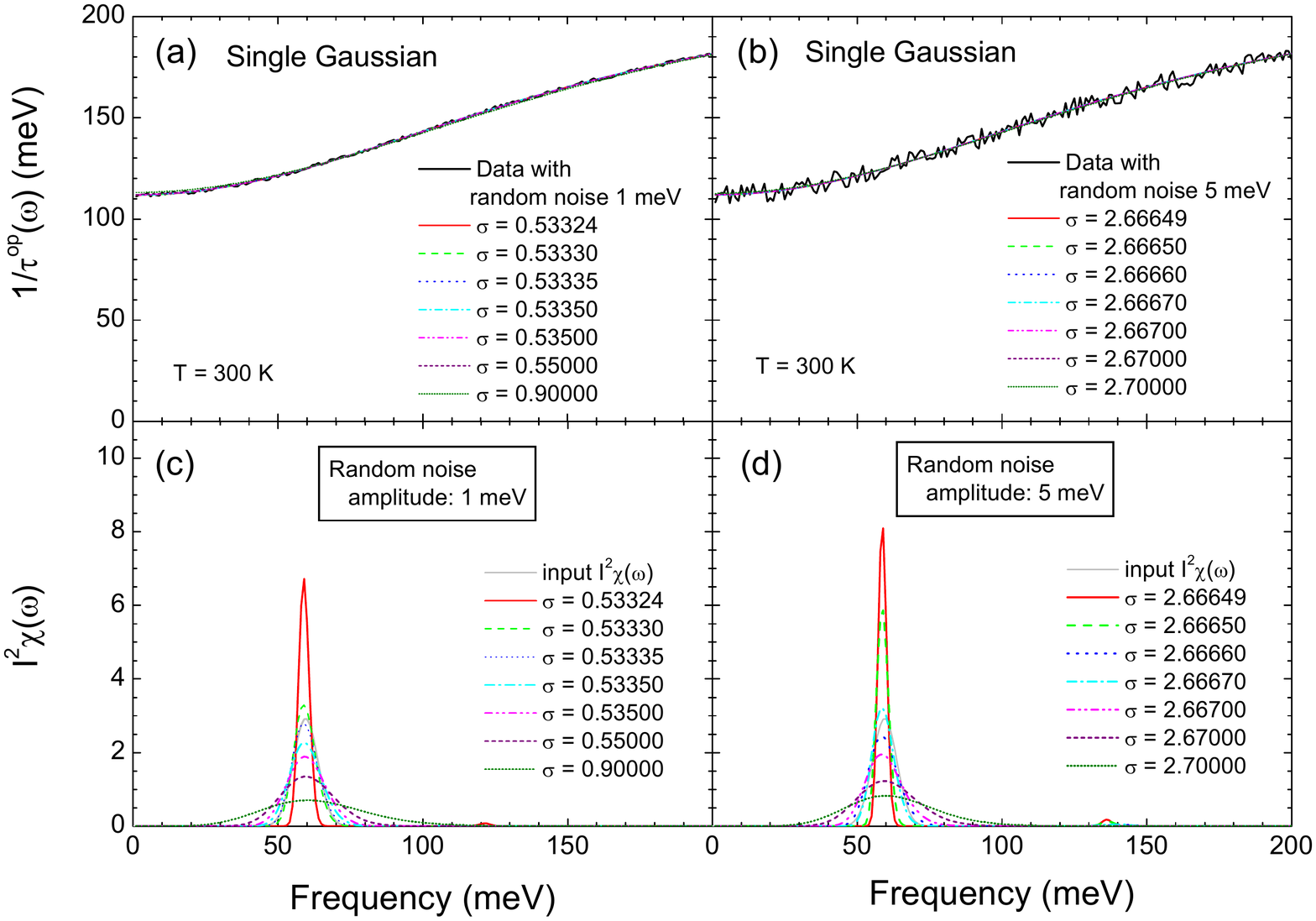}}%
  \vspace*{-0.6cm}%
 \label{fig2a}
\end{figure}

\begin{figure}[t]
  \vspace*{-0.3 cm}%
  \centerline{\includegraphics[width= 7.0 in]{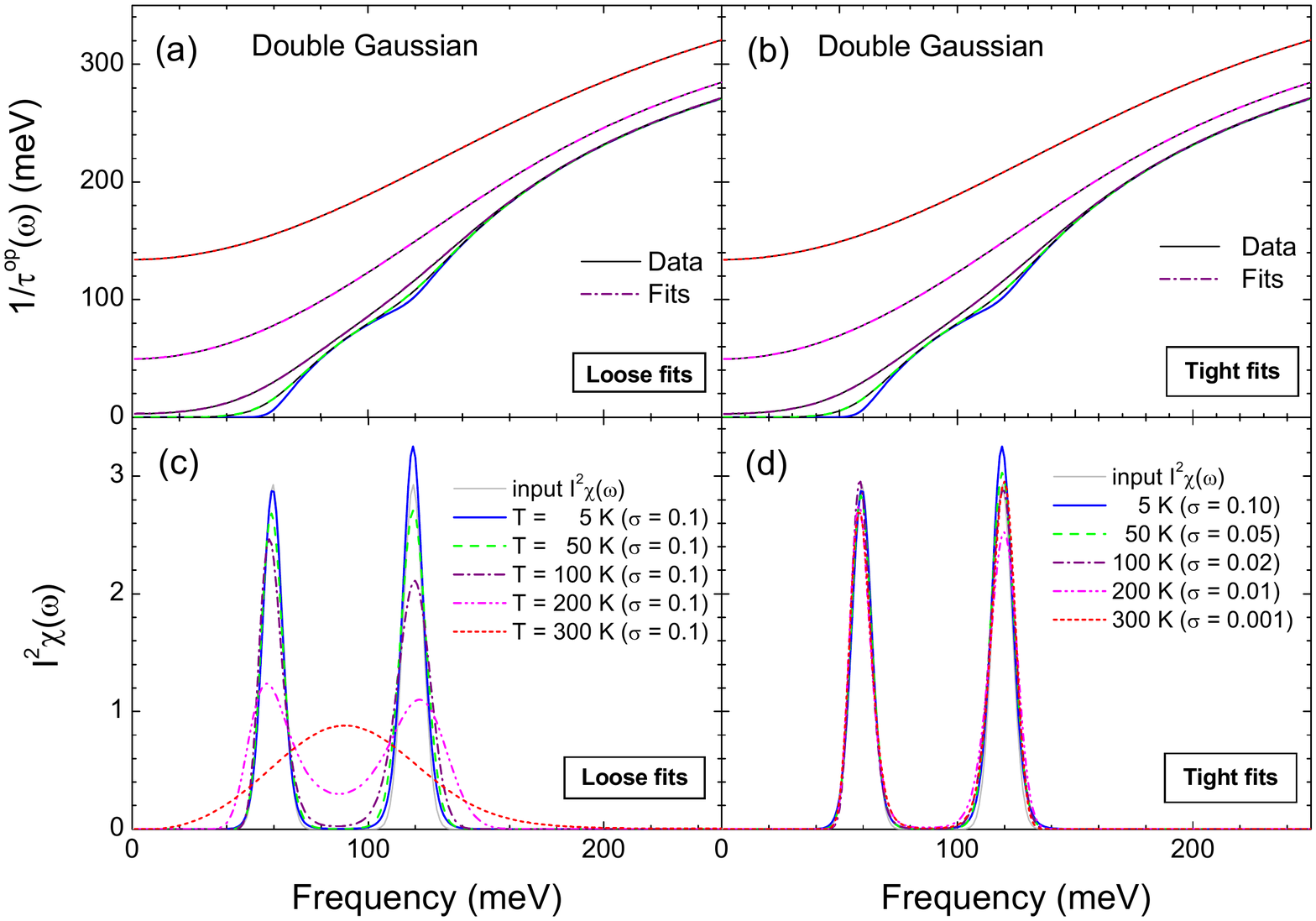}}%
  \vspace*{-0.6cm}%
 \label{fig2}
\end{figure}

\begin{figure}[t]
  \vspace*{-0.3 cm}%
  \centerline{\includegraphics[width= 7.0 in]{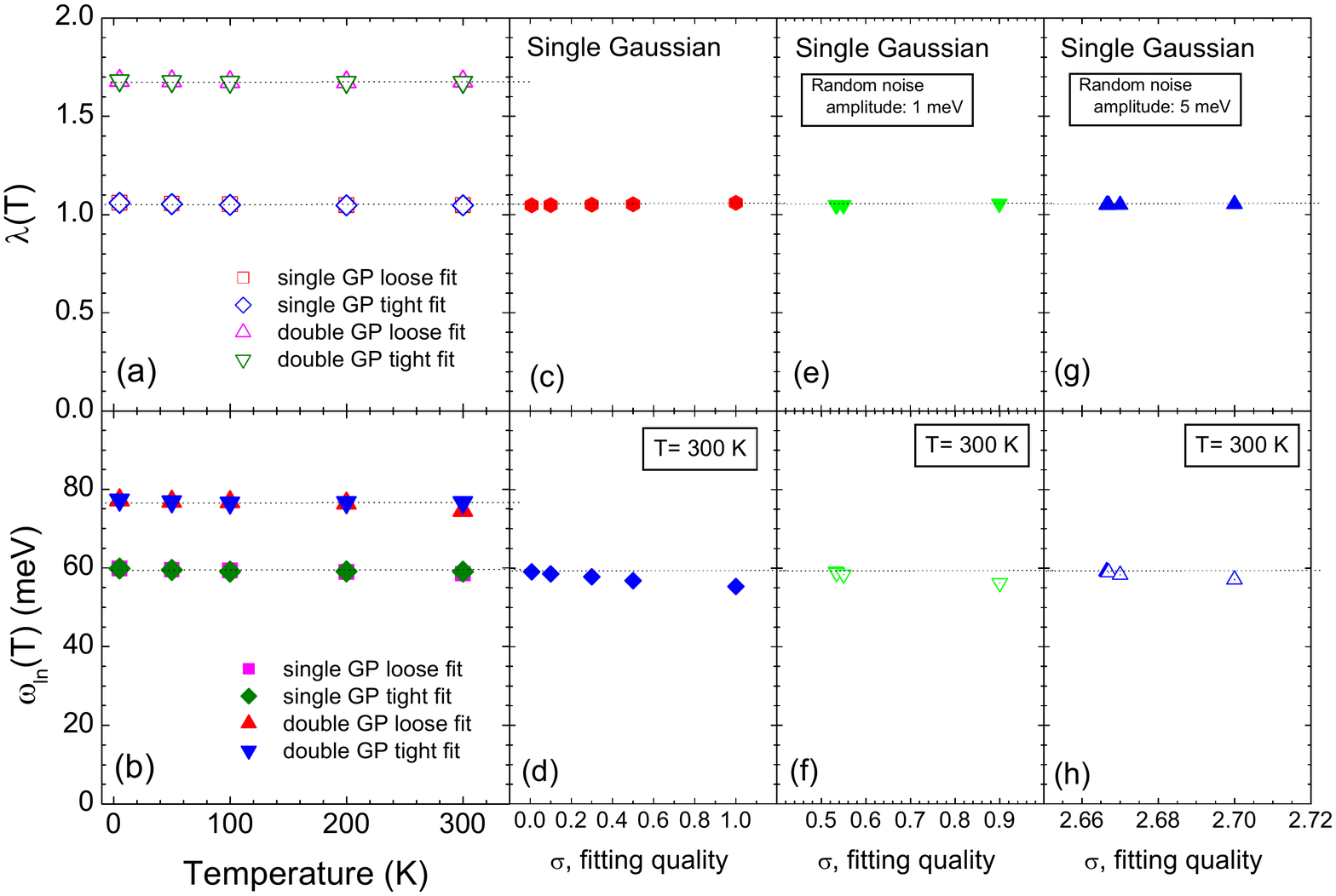}}%
  \vspace*{-0.6cm}%
 \label{fig3}
\end{figure}

\begin{figure}[t]
  \vspace*{-0.3 cm}%
  \centerline{\includegraphics[width= 7.0 in]{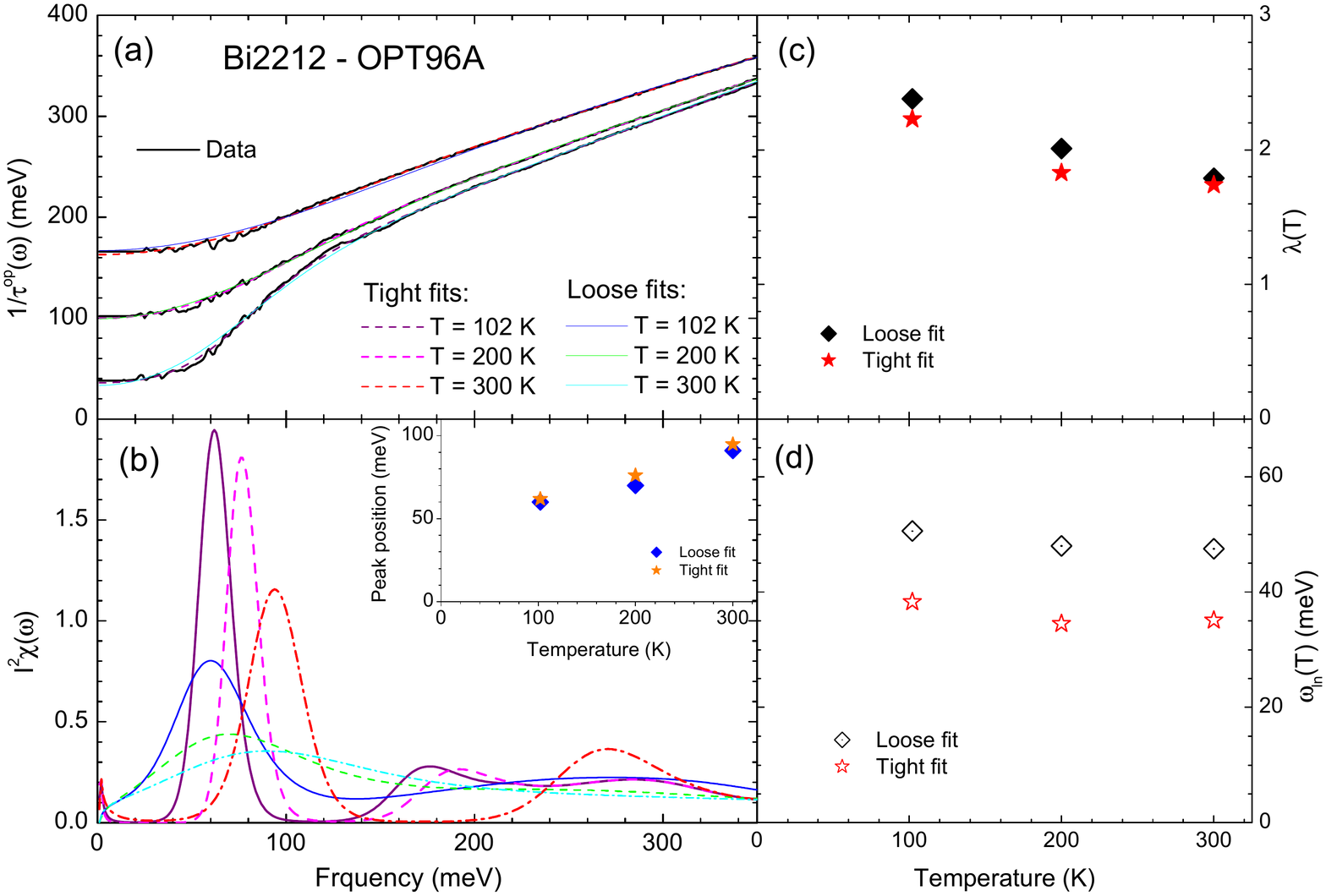}}%
  \vspace*{-0.6cm}%
 \label{fig4}
\end{figure}

\begin{figure}[t]
  \vspace*{-0.3 cm}%
  \centerline{\includegraphics[width= 7.0 in]{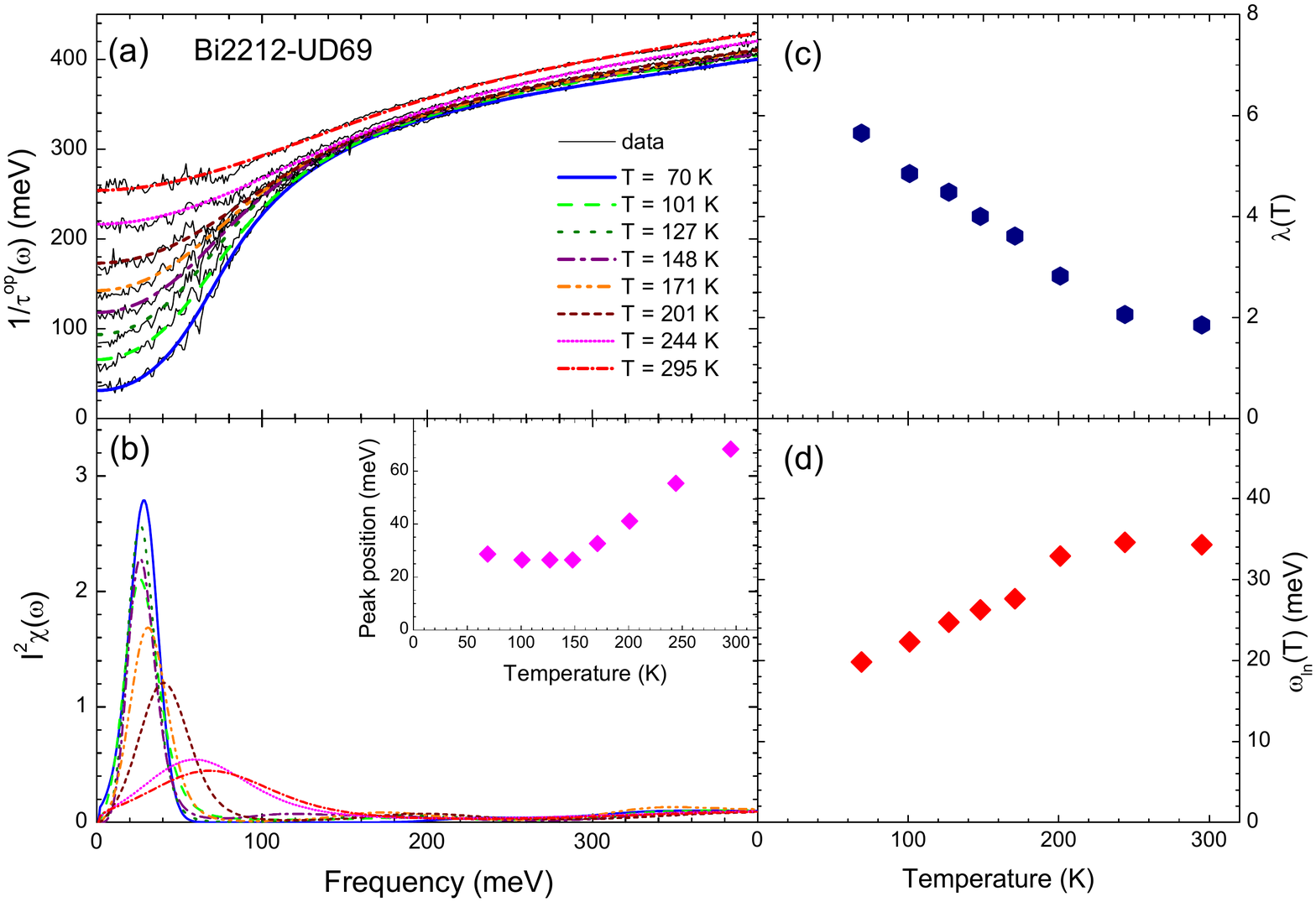}}%
  \vspace*{-0.6cm}%
 \label{fig5}
\end{figure}

\begin{figure}[t]
  \vspace*{-0.3 cm}%
  \centerline{\includegraphics[width= 5.0 in]{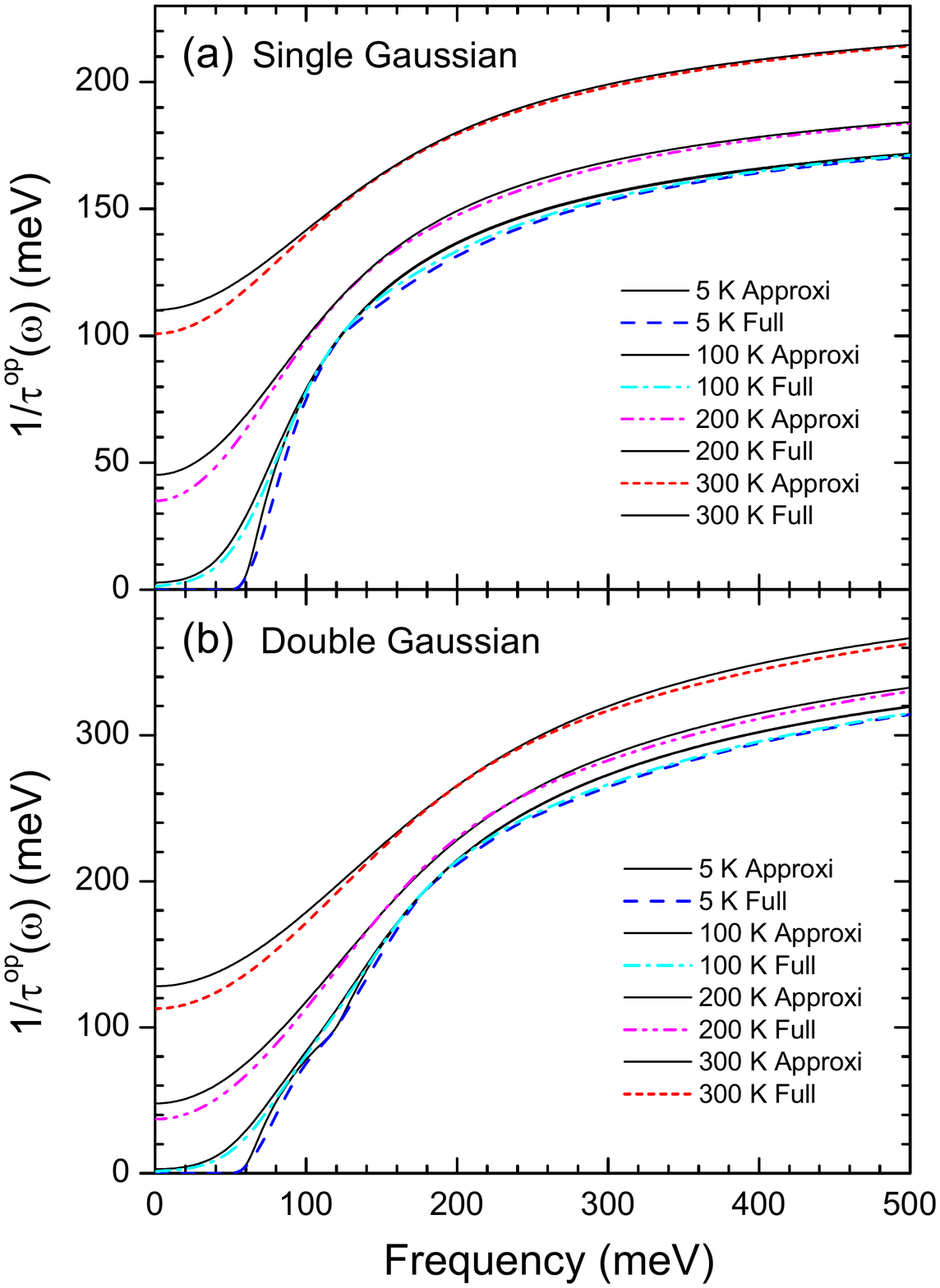}}%
  \vspace*{-0.6cm}%
 \label{fig6}
\end{figure}

\begin{figure}[t]
  \vspace*{-0.3 cm}%
  \centerline{\includegraphics[width= 7.0 in]{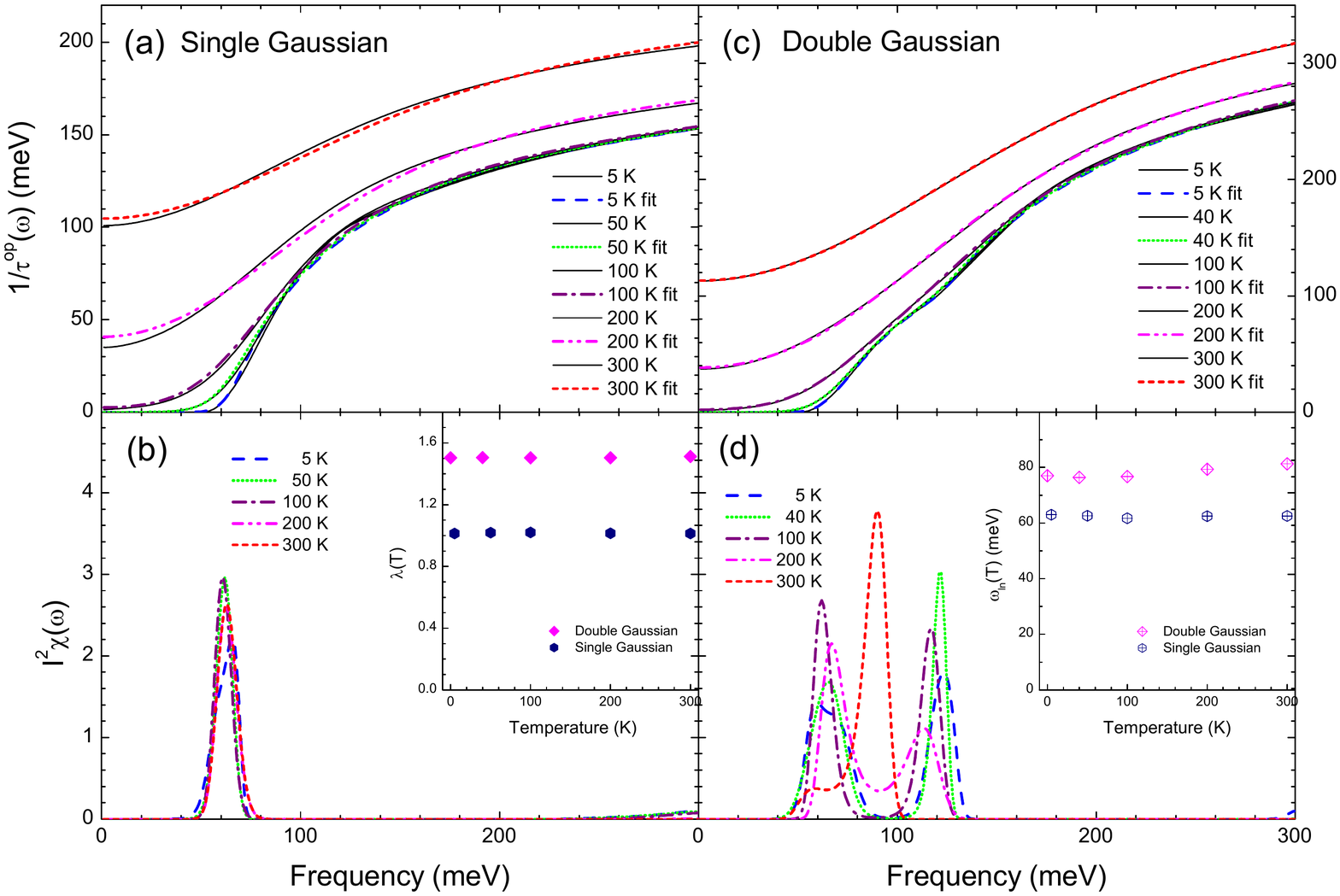}}%
  \vspace*{-0.6cm}%
 \label{fig7}
\end{figure}

\end{document}